\documentclass[showpacs,twocolumn,floatfix,prl]{revtex4}
\usepackage{graphicx}
\usepackage{amsfonts}
\usepackage{amsmath}
\usepackage{amssymb}

\begin{document}

\title{Thermal equilibration between two quantum systems}
\author{A.~V. Ponomarev}
\author{S. Denisov}
\author{P. H\"{a}nggi}
\affiliation{Institute of Physics, University of Augsburg,
Universit\"{a}tstr.~1, D-86159 Augsburg}
\date{\today}

\begin{abstract}
Two identical {\it finite} quantum systems prepared initially at
different temperatures, isolated from the environment, and subsequently
brought into contact are demonstrated to relax towards Gibbs-like
quasi-equilibrium states with a common temperature and small fluctuations
around the time-averaged expectation values of generic observables.
The temporal thermalization process proceeds via a chain of intermediate
Gibbs-like states. We specify the conditions under which this scenario
occurs and corroborate the quantum equilibration with two different models.
\end{abstract}

\maketitle

The derivation of thermodynamic phenomena from deterministic
time-reversible dynamics constitutes one of the primary goals of
physics. This long-standing  conundrum has sparked
recently a new wave of activity in the quantum domain, where
current studies of the objective  follow essentially two tracks. The
first one, pioneered by Schr\"{o}dinger \cite{sred1}, leads to an
understanding of canonical thermalization
when the system of interest is coupled to a much larger system, a quantum
Giant \cite{tasaki,gold,reimann, popescu1, popescu2}.
The studies along the second track explore the ``microcanonical''
thermalization within a single isolated quantum system \cite{nn,
deuch, sred, peres1, gaspard, zanardi1}, and employ exact numerical diagonalization of
many-body models \cite{rigol1,rigol3}.

Here, we focus  on a different route by elucidating the process
of \textit{mutual} equilibration between two finite quantum
``peers", prepared initially at different temperatures and then set
into a contact. We consider two systems, ${\rm A}$ and ${\rm B}$,
that are identical, in the sense that they have identical
Hamiltonians, $H_A=H_B \equiv H_S$. The Hamiltonian $H_S$ has
$\mathcal{N}_S$ non-degenerate energy levels, $\{\epsilon_{k}\}$,
$k=1,..., \mathcal{N}_S$, i.e., $H_S\vert \phi_k \rangle =
\epsilon_k \vert \phi_k \rangle$, with eigenstates $\{\vert \phi_k
\rangle\}$. The systems interact through a contact, which allows 
only for energy transfer without  exchange of particles. The
Hamiltonian of the composite bipartite system thus reads
\begin{equation}
\label{eq:setup} H^\lambda = H_A \otimes \mathbf{1}_B
+ \mathbf{1}_A\otimes H_B + \lambda H^{\rm int},
\end{equation}
with $\lambda$ being a dimensionless coupling constant.
The interaction Hamiltonian, $H^{\rm int}= Y_A \otimes Y_B$,
with operators $Y_A = Y_B \equiv Y$, is invariant under
permutation $A \leftrightarrow B$ and does not commute with the
Hamiltonian $H_S$ \cite{fnote}.

We denote the energy eigenvalues and the corresponding eigenstates
of the Hamiltonian $H^{\lambda}$ by $\{E^{\lambda}_n\}$ and
$\{\vert\psi^\lambda_n\rangle\}$, respectively. The quantities of
interest, i.e. the energy level populations, $p_k^{A}(t)$ and
$p_k^{B}(t)$, can conveniently be calculated by using the product basis,
$\vert \psi^0_{n(k,j)} \rangle = \vert \phi_k
\rangle\otimes\vert\phi_j\rangle$, which is also the eigenbasis of
the composite system for $\lambda = 0$. We label the
energies $E^{0}_n$ according to their decomposition into the sum of
the single system energies, $E_{n(k,j)}^0=\epsilon_k+\epsilon_j=E_{n(j,k)}$. To
shorten notations, we shall  use either $n$ or $kj$ instead of
$n(k,j)$. While combinations $k=j$ produce the non-degenerate energy
levels, $E_{kk}^0=2\epsilon_k$, each two levels related by the
permutation of indices $k \leftrightarrow j$, with $k\neq j$, are
doubly degenerate; i.e. $E_{kj}^0 = E_{jk}^0$. The transformation
from the product basis $\vert \psi_n^0 \rangle$ to the eigenbasis at
a certain interaction strength $\lambda > 0$, $\vert \psi_n^\lambda
\rangle$, is given by the matrix $\mathbf{\Lambda}$, with the
elements $\Lambda_{n,m} = \langle \psi_{m}^0 \vert
\psi_n^\lambda \rangle$. Throughout this work we further assume for
the Hamiltonian (\ref{eq:setup}) with $\lambda\neq 0$ both the non-degeneracy,
$E_n^\lambda \neq E_m^\lambda$ for $n\neq m$, and the ``non-degenerate
energy gap condition" \cite{tasaki, popescu1, popescu2, peres1},
meaning that non-zero energy differences $E^{\lambda}_n - E^{\lambda}_m$ and
$E^{\lambda}_s - E^{\lambda}_w$ are not equal, apart from the
trivial case $s=n,$ $w=m$.

The energy level populations $p^{A(B)}_k(t)$ for system $A$($B$) are given
by the partial trace over system $B$($A$) of the composite system density
matrix $\varrho(t)$; for example, $p^A_k(t)=\sum_{j}\varrho_{kj,kj}(t)$,
where $\varrho(t)$ is expressed in the product basis. In
the case of canonical initial states, where only diagonal density
matrix elements are initially non-zero, their evolution can be
described by the linear map
\begin{equation}
\label{eq:evolution} \varrho_{n,n}(t) =
\sum_{m}\left\vert U_{n,m}^\lambda(t) \right\vert^2 \varrho_{m,
m}(0),
\end{equation}
where $U_{n,m}^\lambda(t)=\sum_{l}
e^{-iE_{l}^\lambda t/\hbar} \Lambda_{l,n}^*\Lambda_{l,m}$. It is
apparent that all necessary information is encoded in the
energy spectrum $\{E_n^\lambda\}$ and in the transformation matrix
$\mathbf{\Lambda}$.

For any choice of the system initial states, $\varrho^{A}(0)$ and
$\varrho^{B}(0)$, the mutual equilibration is guaranteed (in a sense
detailed below) as long as
the non-degenerate energy gap condition holds. Due to the parity $A
\leftrightarrow B$ all eigenstates of the Hamiltonian $H^{\lambda}$
are either symmetric, $\vert \psi_{kj}^\lambda \rangle$ $= \vert
\psi_{jk}^\lambda \rangle$, or antisymmetric $\vert
\psi_{kj}^\lambda \rangle$ $= - \vert \psi_{jk}^\lambda \rangle$.
Therefore, for every  eigenstate of the composite system, expectation
values for any local observable $\boldsymbol{O}$ (energy, level populations, etc.),
associated with one quantum peer only, would be the same for the second
peer, $\mathit{O}^A = \mathit{O}^B$. Having the total system
prepared at time $t = 0$ in a product state $\varrho(0)=\varrho^{\rm
A}(0) \otimes \varrho^{\rm B}(0)$, we turn on the interaction by
setting $\lambda > 0$. Then, after some characteristic relaxation time
$\tau_{rel}$, the system is expected to reach  quasi-equilibrium,
where all diagonal elements of the two subsystem reduced density
matrices obey the relation $\varrho_{kk}^{\rm A}(t) \simeq \varrho_{kk}^{\rm B}(t)$ \cite{fnote1}. 
The respective total equilibrium system energies can be evaluated from the condition of
energy  conservation (assuming a diminutive interaction
energy),
\begin{eqnarray}\label{eq:equp}
E_{eq}^{A, B} \simeq \frac{1}{2}[E^A(0)+E^B(0)],
\end{eqnarray}
where $E^S=\sum_{k} \epsilon_k \varrho_{k,k}^S$ with $S = A~\mathrm{or}~B$. This
is not a genuine equilibrium, since the populations still evolve in time
\cite{kamal}, but their recurrences occur on time scale $\tau_{rec}$ which
is larger than any relevant time scale \cite{peres2,zanardi1}.

To gain an analytical insight, we start out from the limiting case
in which the transformation matrix takes on a
simple form: Any infinitesimally small interaction, $\lambda
\rightarrow 0$, will lift the two-fold degeneracy,
$E^{0}_{kj}=E^{0}_{jk}$, yielding the pair of a symmetric and an
antisymmetric eigenstates in the form $\frac{1}{\sqrt{2}}\left(\vert
\psi_{kj}^0\rangle \pm \vert \psi_{jk}^0\rangle\right)$, $k \neq j$.
These eigenstates are non-degenerate and separated by a finite
splitting. The eigenstates whose energies, $E^{0}_{kk}=2\epsilon_k$,
were non-degenerate at $\lambda=0$, are  perturbed marginally only
in this limit. By assuming this so resulting tridiagonal structure
for the transformation matrix $\Lambda_{n,m}$, we find that the
relaxation process leads to the {\it arithmetic-mean quasi-equilibrium
state}, with the corresponding populations reading \cite{sup}
\begin{equation}
\label{eq:result} p_k^{A,B} \simeq \frac{1}{2}\left[p_k^{
A}(0)+p_k^{B}(0)\right].
\end{equation}
This tridiagonal structure is guaranteed to hold as long as each off-diagonal,
non-zero matrix element of the interaction Hamiltonian, $\lambda \vert H^{\rm
int}_{n,m}\vert$, is smaller than the corresponding energy level difference in the
composite system, $\Delta E_{n,m} = \vert E^{0}_{n} - E^{0}_{m} \vert$.

The characteristic feature of the arithmetic-mean equilibration is
that two systems, when initially prepared in canonical states
at different temperatures,
$\varrho^{S}_{can}(T_S)=e^{-H_S/k_\mathcal{B}T_S}/Z_S$, $Z_S={\rm Tr}(e^{-H_S/k_\mathcal{B}T_S})$,
with the diagonal elements
\begin{eqnarray}
\label{eq:initial} \varrho_{k,k}^{S} (T_S) \equiv
p_k^{S} = \frac{1}{Z_S} e^{-\epsilon_k/k_\mathcal{B}
T_S},
\end{eqnarray}
where $k_\mathcal{B}$ is the Boltzmann constant,
do relax to states with the same mean energy, but their energy level populations,
Eq.~(\ref{eq:result}), are no longer Gibbs-like.
In order to deviate from the limit in Eq.~(\ref{eq:result}) the
transformation matrix $\mathbf{\Lambda}$ needs to acquire a more complex
structure. This is achieved by cranking up the interaction strength
between the two systems. Provided that there occurs a
sufficiently large number of non-vanishing off-diagonal elements, $H^{\rm
int}_{n,m}$, increasing the strength of interaction, but still remaining
within the \textit{weak coupling limit}
\begin{eqnarray}
\label{eq:condition} \lambda (\epsilon_{\mathcal{N}}^{\rm int}-\epsilon_1^{\rm int}) \ll \epsilon_{N_S}-\epsilon_1,
\end{eqnarray}
wherein $\{\epsilon_n^{\rm int}\}$ is the spectrum of the interaction Hamiltonian $H^{\rm int}$,
then yields interaction blocks in the matrix $\Lambda_{n,m}$
larger than those  $2 \times 2$ blocks. We expect that the presence of
a more complex block structure ensures the evolution of
canonical initial states, $\varrho^{A}_{can}(T_A)$ and $\varrho^{B}_{can}(T_B)$,
towards a common Gibbs-like equilibrium $\varrho^{A, B}(T_F)$, meaning
that the corresponding {\it diagonal elements} are given by the relation
(\ref{eq:initial}) with the common temperature $T_F$. The {\it `equilibrium'
temperature} $T_F$ can be evaluated from Eq.~(\ref{eq:equp}), to yield with Eq.~(\ref{eq:initial}):
\begin{eqnarray}\label{eq:temp_eq}
\sum_k\varepsilon_k\frac{e^{-\frac{\varepsilon_k}{k_\mathcal{B}T_F}}}{Z_F} = \frac{1}{2}\sum_k\varepsilon_k\left[\frac{e^{-\frac{\varepsilon_k}{k_\mathcal{B}T_A}}}{Z_A} + \frac{e^{-\frac{\varepsilon_k}{k_\mathcal{B}T_B}}}{Z_B}\right].
\end{eqnarray}

\begin{figure}[t]
\includegraphics[width=0.45\textwidth]{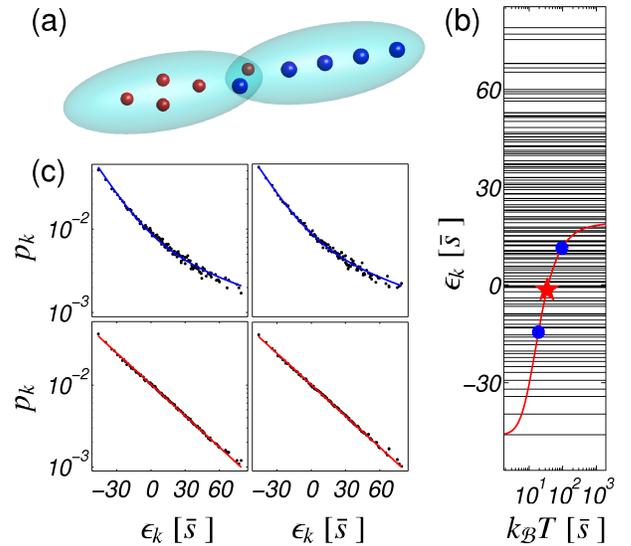}
\caption{(color online) (a) A system of bosons confined into two
overlapping confinements is analyzed with the Bose-Hubbrad model.
(b) Energy spectrum of a single system. The (red) line
displays the dependence of the system mean energy, i.e., $E^S=\sum_{k} \epsilon_k
e^{-\epsilon_k/k_\mathcal{B} T}/Z_S$, on temperature $T$. The initial
temperatures of the `hot' system, $k_\mathcal{B} T_A/\bar{s}= 94.91$,
and the `cold' system, $k_\mathcal{B} T_B/\bar{s}= 18.98$, are
indicated by the (blue) dots. The equilibrium temperature, $k_\mathcal{B}T_F/\bar{s}=33.92$,
calculated by using the total energy conservation, Eq.~(\ref{eq:temp_eq}),
is indicated by the (red) star. (c) Instantaneous `equilibrium' energy
level populations for systems $A$ (left column) and $B$ (right column),
in the regime of arithmetic-mean (top) and thermal (bottom) equilibrations.
The arithmetic-mean populations are depicted by the top (blue) solid
lines, and the canonical populations for the temperature $T_F$ by the
bottom (red) lines. The natural energy unit, $\bar{s}$, is given by
the mean energy level spacing of the single system,
$\bar{s}=(\epsilon_{\mathcal{N}_{S}}-\epsilon_1)/(\mathcal{N}_S-1)$.
The similar behavior is demonstrated by the second model, see
supplementary material for further model details.}
\label{fig1}
\end{figure}

We numerically validate our prediction by using two types of
quantum models. Within the Bose-Hubbard model we consider
the system consisting of $N=5$ on-site interacting bosons
on a one-dimensional lattice, with $L = 5$ sites and
hard-wall boundaries. This results
in $\mathcal{N}_S=\frac{(L+N-1)!}{(L-1)!N!}=126$ energy
levels in each single system, and $\mathcal{N} = \mathcal{N}_S
\times \mathcal{N}_S = 15,876$ levels in the composite system \cite{sup}.
Figure~1(a) depicts the setup, which assumes that the two systems
overlap only by one site, where the bosons from the different
confinements do interact. We also corroborated our findings with a
randomly synthesized model, for which the Hamiltonian $H_S$ and the
interaction operator $Y$ are independently sampled from a finite-dimensional
Gaussian Orthogonal Ensemble (GOE) of random matrices \cite{sup}. 
In contrast to the former many-body interacting boson model,
where the interaction is strictly local, here the interaction is
now acting globally, interweaving systems $A$ and $B$.

For both models we find solutions that are based on the exact
diagonalization of the corresponding bipartite Hamiltonians. Our
main results are depicted in Fig.~\ref{fig1}. Upon increasing
the coupling constant $\lambda$ within the weak coupling limit,
Eq.~(\ref{eq:condition}), we detect a crossover from the
arithmetic-mean quasi-equilibrium populations, Eq.~(\ref{eq:result}),
towards the canonical populations, Eq.~(\ref{eq:initial}) with $T_S=T_F$.

\begin{figure}[t]
\center
\includegraphics[width=0.45\textwidth]{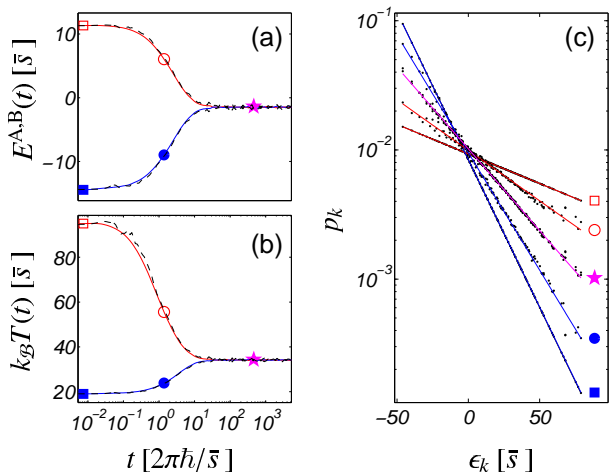}
\caption{(color online) Relaxation pathways for the model
depicted in Fig.~1(a). Both systems are initially prepared in
canonical states (solid lines) and in pure states randomly sampled
from the corresponding ensembles of typical states, Eq.~(\ref{eq:typical}) 
(dashed lines). (a) The evolution of the mean energies and (b) the 
corresponding temperatures $T(t)$ of the `hot' system ${A}$
and the `cold' system ${B}$ are shown by the upper (red) and lower (blue)
lines, respectively. (c) The energy level populations of both
systems are displayed at different moments of time (dots), 
marked by the corresponding symbols in (a, b). The lines correspond to the
canonical populations, Eq.~(\ref{eq:initial}), at the temperatures evaluated from 
the temporal values of mean system energies (see Fig.~1(b)).} \label{fig2}
\end{figure}

An intriguing question is how the quantum equilibration unfolds in time.
Figure~2 displays our finding that
equilibration proceeds along a quasistatic pathway: the relaxation
of an initial canonical state abides a sequence of time-dependent
Gibbs-like states with time-dependent temperatures $T(t)$,
intermediate between the initial temperature $T_{A (B)}$, to reach a
common, final temperature $T_F$. This observed persistence of
Gibbs shape is remarkable indeed. The only relevant result we
could find in this context is that of thermal relaxation dynamics
of a stylized model \cite{shuler}.

\begin{figure}[t]
\center
\includegraphics[width=0.45\textwidth]{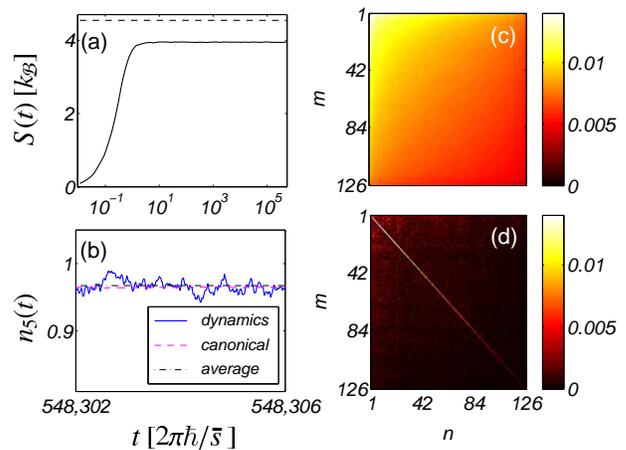}
\caption{(color online) (a) Von Neumann entropy of a single
system {\it vs.} time for the model shown in Fig.~1(a).
Both systems are initially prepared in pure states randomly sampled
from the ensembles of typical states, Eq.~(\ref{eq:typical}). The dashed
line indicates the entropy of the Gibbs state at the temperature
$T_F$. (b) The population dynamics $n_5(t)$ of the fifth-site
(i.e. the site making the thermal contact) for the subsystem $A$ is
compared with the corresponding canonical value at the temperature $T_F$.
Note that the time average of $n_5(t)$, $0.9673$, differs from
its canonical value, $0.9651$, by $0.3\%$ only. (c) The absolute values of 
reduced density matrix elements $\langle\phi_m\vert\varrho^{A}(t)\vert\phi_n\rangle$ 
at $t=0$, and (d) after equilibration.}
\label{fig3}
\end{figure}

We next  consider the case with an initial preparation given by
pure  states. Reproducibility of quantum thermal processes with a
\textit{single} `typical' state \cite{typical1} carries importance
in view of the foundations of statistical physics
\cite{nn, typical2} and many-body quantum calculations \cite{white}.
We employ here typical states constructed as the sums over eigenstates \cite{typical1}; 
i.e., we use
\begin{equation}
\label{eq:typical} \vert \psi^{S}_{T_{S}}(0) \rangle  =
\frac{1}{\sqrt{Z_{S}}}\sum_{k} e^{i \theta_{k}^S}
e^{-\epsilon_k/2 k_\mathcal{B} T_{S}}\vert \phi_{k}\rangle.
\end{equation}
The ensemble of typical states is defined by the uniform measure on
the torus $\theta_{1}^S \otimes \theta_{2}^S...\otimes \theta_{N}^S$,
$\theta_{k}^S \in [0, 2\pi]$. The results shown in Fig.~2(a,~b) by the
dashed lines confirm our expectation: A \textit{single}, randomly
sampled, initial product wave function $\vert\psi_{T_{A}}^{A}(0) \rangle
\otimes \vert \psi_{T_{B}}^{B}(0)\rangle$ follows the equilibration
pathway for canonical initial states with good accuracy. Both systems,
${A}$ and ${B}$,  are prepared initially in pure states, implying vanishing
von Neumann entropies  $S_{A, B}(t) = - k_\mathcal{B} {\rm Tr}[\varrho^{A, B}(t) \ln \varrho^{A, B}(t)]$, i. e., $S_{
A}(0)=S_{B}(0)=0$. The isolated composite system remains in a pure
state forever, and thus $S_{{A}\otimes{ B}}(t)\equiv 0$. This, however,
is no longer so for subsystem entropies $S_{ A}(t)$ and
$S_{B}(t)$, which start to grow. From the triangle inequality it
follows that $S_{A}(t)=S_{B}(t) \equiv S(t)$. The entropy $S(t)$ is
a measure for entanglement between the subsystems \cite{ent}: its
monotonic growth  thus indicates that the equilibration process
entangles the quantum peers, see Fig.~3(a).

The systems cannot rigorously reach canonical equilibrium; therefore,
the entropy $S(t)$ saturates to the value below the entropy of the
Gibbs state at temperature $T_F$. The resulting equilibrium system
density matrices, $\varrho^A(t)$ and $\varrho^B(t)$, remain
nonstationary and possess both diagonal and off-diagonal elements
evolving in time. Following the recipe from Ref.~\cite{popescu2},
the deviation from the canonical state is estimated by using
the trace-norm distance $\mathcal{D} =
\mathrm{Tr}\langle\vert\varrho^S(t) -
\bar{\varrho}^S \vert\rangle_{t}/2$, where the bar denotes the time average $\langle \dots \rangle_{t}$. This
quantity is limited from above \cite{popescu2}, so that from Eq.~(8)
in \cite{popescu2} we find that $\mathcal{D} \leq 0.6 $ in our
case. From our numerics we obtain $\mathcal{D} \simeq 0.43$.

For an operator $\boldsymbol{O}$, which is non-diagonal in the eigenbasis of the Hamiltonian $H_S$,
the presence of the off-diagonal elements in the system density matrices will produce additional
fluctuations around the average value $\mathit{\bar{O}}^S = \mathrm{Tr}(\bar{\varrho}^S\boldsymbol{O})$.
Moreover, some of the off-diagonal elements may possess non-zero time averages. This might cause a
constant shift of the observable averaged value $\mathit{\bar{O}}$
from its canonical value, $\delta \mathit{O}^S = \mathit{\bar{O}}^S -
\mathrm{Tr}[\bar{\varrho}_{can}^S(T_F)\boldsymbol{O}]$. However for highly non-sparse
patterns of non-zero off-diagonal elements $\mathit{O}_{kl}^S$ and $\varrho^{S}_{kl}$,
we may expect that the respective fluctuations of the
expectation value $\mathit{O}^S(t)$ will be suppressed, exhibiting
{\it dynamical typicality} \cite{gemmer1}.
Even for a system as small as ours, with $\mathcal{N}_S = 126$ states,
this mechanism works surprisingly well, see Fig.~3(b).

Thermal quantum relaxation within an isolated composite quantum
system is a deterministic process, and produces an output in the
form of a Gibbs-like equilibrium, with diagonal elements which are almost canonical,
for the initial preparation, Eq.~(\ref{eq:initial}), and also for initial `typical' pure states,
Eq.~(\ref{eq:typical}). An arbitrary choice of the initial state of the composite
system $H^\lambda$ does not guarantee relaxation towards Gibbs-like quasi-equilibrium
states for its halves. Also the state of the composite system after relaxation
is far from being Gibbs-like due to strong entanglement between its halves.
Moreover, in order to render the thermodynamical relaxation of quantum peers,
two necessary conditions need to be fulfilled, namely, (i) the interaction
is restricted to the validity range of Eq.~(\ref{eq:condition}), and (ii) the total composite
system obeys the parity $A \leftrightarrow B$. A natural question then is: What will
happen if either of the conditions (i) or (ii) is violated?
For (i) the systems will nevertheless equilibrate
even with the  interaction strength  set beyond the
weak-coupling limit. The  corresponding `equilibrium' state,
however,  no longer assumes a Gibbs-like structure. The part
(ii) with non-identical systems $A$ and $B$ is more intricate.
Although it is still possible to obtain thermal relaxation between
two different systems (see \cite{sup}), the mismatch of system spectra and
their relatively small sizes necessitates a much larger
system-system coupling constant $\lambda$ \cite{sup}. The resolution of this problem
demands systems of much larger sizes, and, therefore, lies outside
the exact diagonalization scheme employed here.

The quasistatic character of the thermal relaxation allows for the tuning of one
of the two quantum peers to a Gibbs-like state at any temperature between initial
temperature values, $T_A$ and $T_B$, thus serving as an alternative protocol for
the preparation of thermal states of quantum systems \cite{boixo}. The state-of-the-art
experiments with ultracold atoms provide the natural playground for exploration
of the thermal relaxation between two different species of atoms \cite{inguscio}.

This work is supported by the DFG grant HA1517/31-2 and by the ``Nanosystems Initiative Munich'' (NIM).


\begin{thebibliography}{30}
\bibitem{fnote} One could consider a more general form of the interaction Hamiltonian
with the Schmidt rank $R>1$, i.e. $H^{\rm int}=\sum_{r=1}^R \lambda_r Y_A^r\otimes Y_B^r$.
This however would not affect the generality of our further analysis.
\end{thebibliography}
\end{document}

\title{Supplementary material for \\`Thermal equilibration between two quantum systems'}
\author{A. V. Ponomarev}
\author{S. Denisov}
\author{P. H\"{a}nggi}
\affiliation{Institute of Physics, University of Augsburg,
Universit\"{a}tstrasse~1, D-86159 Augsburg}

\pacs{03.65.Aa, 05.30.-d, 05.40.-a, 05.70.Ln, 67.85.-d
}

\maketitle

\begin{figure*}[t]
\leavevmode\includegraphics[width=0.95\hsize]{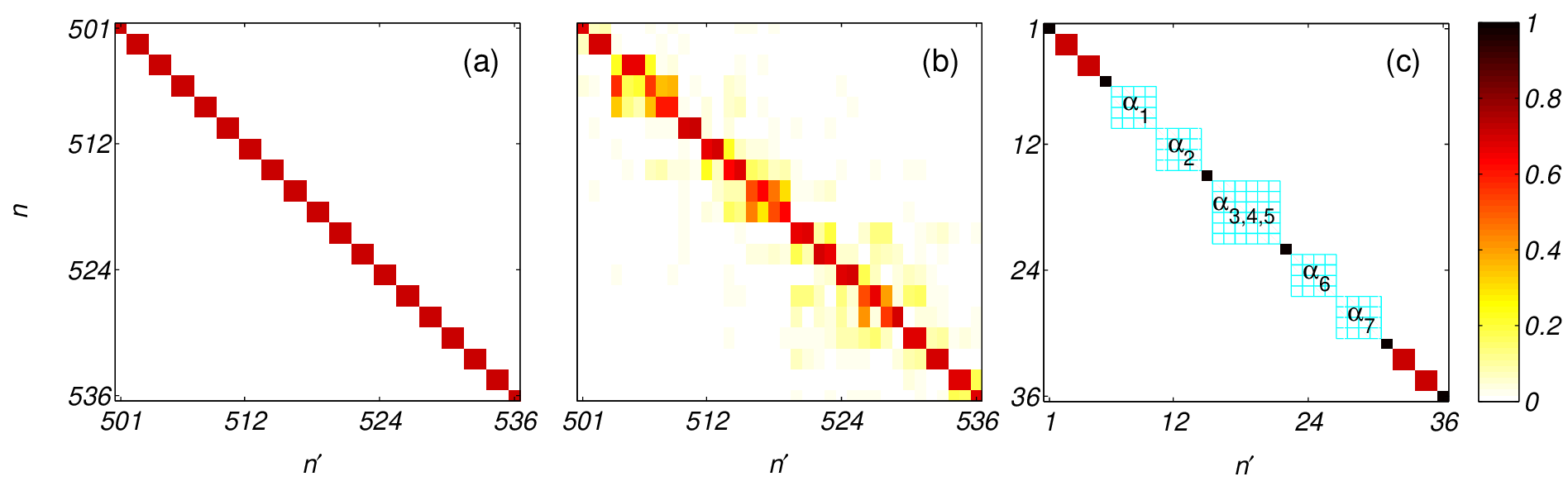}
\caption[Transformation matrix] {{\bfseries Structure of transformation matrix.}
(a) The part of transformation matrix, $\Lambda_{n,n'}$, for the GOE model of quantum peers in the limit of arithmetic-mean equilibration. (b) The same part of transformation matrix beyond the arithmetic-mean limit. (c) The model of the $36 \times 36$ transformation matrix with four $2 \times 2$ blocks, four $4 \times 4$ blocks, one $6 \times 6$ block, and the seven corresponding free parameters, $\alpha_i$, $i=1,...,7$. Only the absolute values of transformation matrix elements are shown.}
\label{fig1s}
\end{figure*}

\section{Arithmetic-mean equilibration}

The arithmetic-mean equilibration occurs in the limit of weak
coupling between the systems, $\lambda\{ H^{\rm int}\} \ll
\{\Delta E \}$, where the transformation matrix
$\Lambda_{n,m}=\langle \psi_{m}^0 \vert \psi_n^\lambda \rangle$
effectively takes on the tridiagonal form,  see Fig.~\ref{fig1s}(a),
with non-zero entries:
\begin{equation}
\label{eq:Lambda_S}
\Lambda_{n(k,j),m(k',j')}=  \left\{
\begin{array}{cl}
1, & j=j'=k=k'; \\
1/\sqrt{2},& j=j'\neq k=k';\\
\chi_{kj}/\sqrt{2}, & j=k'\neq k=j',
\end{array}
\right.
\end{equation}
where $\chi_{kj}={\rm sgn}(k-j)$. Hence the dynamics of the composite system is
governed by the tridiagonal evolution operator:
\begin{eqnarray}
\label{eq:evol_opr}
U_{n(k,j),m(k',j')}^\lambda(t)=\sum_{m} e^{-iE_{m}^\lambda t/\hbar} \Lambda_{m,n}^*\Lambda_{m,n'}\nonumber \\
=\left\{
\begin{array}{cl}
1, & j=j'=k=k'; \\
e^{i\Omega_{kj}^\lambda}\cos\left(\frac{\omega_{kj}^\lambda t}{2}\right), & j=j'\neq k=k';\\
ie^{i\Omega_{kj}^\lambda}\chi_{k'j'}\sin\left(\frac{\omega_{kj}^\lambda t}{2}\right), & j\neq k, j=k'\neq k=j',
\end{array}
\right.
\end{eqnarray}
where $\Omega_{kj}^\lambda=(E_{kj}^\lambda+E_{jk}^\lambda)/2\hbar$ and $\omega_{kj}^\lambda=(E_{kj}^\lambda-E_{jk}^\lambda)/\hbar$.
For the density matrix of the composite system written in the
product basis, $\vert \psi^0_{n(k,j)} \rangle$, knowledge of its
diagonal elements is sufficient for calculations of the energy level
populations. For example,  for the system ${ A}$ one has
\begin{equation}
\label{eq:popul}
p_k^{ A}(t)  = \sum_{j=1}\varrho_{n(k,j),n(k,j)}(t).
\end{equation}

\subsection{Evolution of initially diagonal states}

For the systems whose initial density matrices are diagonal in
$\{\vert \phi^{0}_n\rangle\}$, $\rho^{
A,B}(0)_{k,k'}=\delta_{k,k'}p^{ A,B}_k(0)$,
the evolution of the diagonal matrix elements for the composite
system density matrix, $p_{n}(t)\equiv\varrho_{n,n}(t)$, reduces
to the linear map: $p_n(t) = \sum_{n'}\left\vert
U_{n,n'}^\lambda(t) \right\vert^2 p_{n'}(0)$. By using the
evolution operator, Eq.~(\ref{eq:evol_opr}), we obtain
\begin{eqnarray}
p_{kj}(t)&=&\frac{1}{2}\left(p_{kj}(0) + p_{jk}(0)\right)\nonumber \\
&&\quad + \frac{1}{2}\cos\left(\omega_{kj}^\lambda t\right)\left(p_{kj}(0) -  p_{jk}(0)\right).
\end{eqnarray}
The energy level populations of one system, e.g. the system ${
A}$ are
\begin{eqnarray}
\label{eq:popul_D}
p_k^{ A}(t) = \frac{1}{2}\left[p_k^{ A}(0)+p_k^{ B}(0)\right] + \frac{1}{2}\sum_j X^{ A}_{kj} \cos\left(\omega_{kj}^\lambda t\right),
\end{eqnarray}
where $X^{ A}_{kj}=p_k^{ A}(0) p_j^{ B}(0) - p_j^{ A}(0) p_k^{ B}(0)$.

In order to gain some physical insight it is useful to explore the following situations:
(i) one of the two systems, for an example system $B$, is initially
localized on a single level, i.e. $p_k^{B}=\delta_{k,k_B}$, and (ii) both the systems,
$A$ and $B$, are initially localized on single levels, i.e. $p_k^{A}=\delta_{k,k_A}$ and
$p_k^{B}=\delta_{k,k_B}$.

In the first case (i), Eq.~(\ref{eq:popul_D}) reduces to
\begin{eqnarray}
\label{eq:popul_case1a}
p_k^A(t) & \nonumber \\
 = & \frac{1}{2} \left\{
\begin{array}{rcll}
p_{k}^A(0)+1&-&\sum_{j\neq k} p_j^A(0)\cos(\omega_{kj}^\lambda t), & k=k_B; \\
p_k^A(0)+ 0&+&p_k^A(0)\cos(\omega_{kk_B}^\lambda t), & k\neq k_B,
\end{array}
\right.
\nonumber
\end{eqnarray}
while for the energy level populations of the system $B$ we obtain,
\begin{eqnarray}
\label{eq:popul_case1b}
p_k^B(t) &\nonumber \\
 = &\frac{1}{2} \left\{
\begin{array}{rcll}
1+p_{k}^A(0)&+&\sum_{j\neq k} p_j^A(0)\cos(\omega_{kj}^\lambda t), & k=k_B; \\
0+p_k^A(0)&-&p_k^A(0) \cos(\omega_{kk_B}^\lambda t), & k\neq k_B.
\end{array}
\right.
\nonumber
\end{eqnarray}
Thus the population of the energy levels with the same index $k\neq k_B$ in the
system $A$ and the system $B$ oscillate coherently, while the populations of energy
levels with $k=k_B$ quasiperiodically fluctuate, for all $k$ around the
arithmetic-mean of the initial populations.

In the second case (ii), the energy level populations of systems $A$ and $B$ read:
\begin{eqnarray}
\label{eq:popul_case2a}
p_k^A(t) & \nonumber \\
 = & \frac{1}{2} \left\{
\begin{array}{rcll}
1 + 0&+& \cos(\omega_{kk_B}^\lambda t), & k=k_A; \\
0 + 1&-&\cos(\omega_{k_Ak}^\lambda t), & k=k_B; \\
0 + 0&,& & k\neq k_A, k\neq k_B;
\end{array}
\right.
\nonumber
\end{eqnarray}
and
\begin{eqnarray}
\label{eq:popul_case2b}
p_k^B(t) & \nonumber \\
 = & \frac{1}{2} \left\{
\begin{array}{rcll}
0 + 1&-&\cos(\omega_{kk_B}^\lambda t), & k= k_A; \\
1 + 0&+&\cos(\omega_{k_Ak}^\lambda t), & k=k_B; \\
0 + 0&,& & k\neq k_A, k\neq k_B.
\end{array}
\right.
\nonumber
\end{eqnarray}
Thus, for $k_A\neq k_B$ each system performs coherent {\it Rabi type} oscillations
between the levels $k_A$ and $k_B$ around the arithmetic-mean of their initial values,
while the rest of the levels remains unpopulated. Evidently, if $k_A=k_B=k_S$ then the 
corresponding frequency $\omega_{k_Sk_S}\equiv 0$, thus both systems stay localized 
forever and there is no dynamics at all.

\subsection{Evolution of initially pure states}

When initially both the systems are assumed to be in pure states,
$\vert\psi^{ A,B}(0)\rangle=\sum_k c_k^{ A,B}\vert \phi_k \rangle$
and the initial state of the composite system is given by their product,
$\vert \psi(0) \rangle =  \vert\psi^{ A}(0) \rangle \otimes \vert\psi^{ B}(0)\rangle$,
so that
\begin{equation}
\label{eq:initial_tot_S}
\vert \psi(0) \rangle =  \sum_n c_{n(k,j)} \vert\psi^0_{n(k,j)}\rangle, \;  c_n=c_k^{ A}c_j^{ B}.
\end{equation}
the diagonal matrix elements of the total density matrix are
\begin{eqnarray}
p_{kj}(t) = \left\vert\langle \psi^0_{kj} \vert \psi(t)\rangle \right\vert^2=  \left\vert\langle \psi^0_{kj}\vert \mathbf{U}^\lambda(t) \vert \psi(0)\rangle \right\vert^2.
\end{eqnarray}
Using $c_{kj}=\sqrt{p_k^{ A} p_j^{ B}}\exp\left[i(\theta_k^A+\theta_j^B)\right]$, where $\theta_k^A$ ($\theta_j^B$), 
the phase entering the initial state of the system $A$ ($B$) (see Eq.~(8), of main manuscript), we end 
up with the the energy level populations of the system $A$:
\begin{eqnarray}
\label{eq:reduced_rho_result_S}
p_k^{ A}(t) &=& \frac{1}{2}\left[p_k^{ A}(0)+p_k^{ B}(0)\right] + \frac{1}{2}\sum_j X^{ A}_{kj} \cos\left(\omega_{kj}^\lambda t\right) \nonumber \\
&&+\frac{1}{2}\sum_j Y^{ A}_{kj} \sin\left(\theta_{kj} -\theta_{jk}\right)\sin\left(\omega_{kj}^\lambda t\right),
\end{eqnarray}
where $Y^{ A}_{kj}=\chi_{jk}\sqrt{p_k^{ A}(0) p_j^{ B}(0)p_j^{ A}(0) p_k^{ B}(0)}/2$ and $\theta_{kj} = \theta_k^A+\theta_j^B$. The only difference from the Eq.~(\ref{eq:popul_D}) is the last sum on the rhs.
The latter is generated by non-zero off-diagonal elements of initial density matrix, $\varrho(0) = \vert
\psi(0)\rangle\langle\psi(0)\vert$.

Apparently, the evolution for diagonal initial states can also be
obtained from Eq.~(\ref{eq:reduced_rho_result_S}) by
averaging over the ensemble of pure states with the different initial
phases, $\theta_{kj}$, and fixed initial populations of the energy levels,
$p_k^{ A,B}(0)$. This would lead to the nullification of the last
sum on the rhs of Eq.~(\ref{eq:reduced_rho_result_S}).

\subsection{Relaxation to the arithmetic-mean of the initial populations}

In most physical situations, the initial energy level populations,
$p_k^{ A,B}(0)$, exhibit a smooth dependence on $k$, thus producing
a significant number of non-zero coefficients $X^{ A, B}_{kj}$.
This is also the case for the canonical states with $k_\mathcal{B} T
\gg \bar{s}$, where $\bar{s}$ denotes the mean level spacing,
$\bar{s}=(\epsilon_{\mathcal{N}_S}-\epsilon_1)/(\mathcal{N}_S-1)$. The
relation (\ref{eq:popul_D}) yields $p_k^{
A}(0)$ at $t=0$.
In the course of time every member of the sum on the  rhs of
Eq.~(\ref{eq:popul_D}) begins acquiring a certain phase.
Since the frequencies $\omega_{kj}^\lambda$  do not commensurate in
general, after the characteristic time, $\tau_{rel} \sim 2\pi/
\omega_{ typ}^\lambda$, where $\omega_{typ}^\lambda$ is the
root mean square of the set $\{ \omega_{kj}^\lambda\}$, we will
obtain a sum of independent random values, almost uniformly
distributed over the interval $[-1,1]$, and weighted with the
coefficients $X^{ A}_{kj}$. Given the initial states with a
substantial number of populated energy levels, the sum looses its initial
coherence completely after the time $\tau_{rel}$ and averages
itself to zero. Hence the relaxation process leads to the
arithmetic-mean equilibration,
\begin{equation}
\label{eq:result} p_k^{ A,B} \simeq \frac{1}{2}\left[p_k^{
A}(0)+p_k^{ B}(0)\right].
\end{equation}

The rhs of Eq.~(\ref{eq:popul_D}) is a quasiperiodic
function \cite{kamal}, therefore it  will repeat itself after some time
$\tau_{rec}$ with any given accuracy $\Delta$, so that $\| p_k^{
A}(t+\mathcal{T}(\Delta))-p_k^{ A}(t)\| \leq \Delta$. Assuming, for example, that the
coefficients $X^{A, B}_{kj}$ are equal, the recurrence time grows exponentially
with $\mathcal{N}_S$: $\tau_{rec}\sim\frac{1}{\mathcal{N}_S^{1/2}\omega_{typ}^\lambda}\Gamma(\mathcal{N}_S+1)(\Delta
\mathcal{N}_S/4\pi)^{-(\mathcal{N}_S-1)/2}$ \cite{peres}. Already for $\mathcal{N}_S=10$ and $\Delta=0.1$ we
find a sharp scale separation between the recurrence and relaxation times,
$\tau_{rec}/\tau_{rel} \simeq 3 \cdot 10^{6}$.

\section{Blocks in the transformation matrix $\Lambda$ as
initiators of thermal equilibration}

Let us denote by $\vert \psi_{s, a}^{\lambda}\rangle$ the pair, the symmetric and the antisymmetric
eigenstates, and by $\vert \psi_{\mathcal{A},(\mathcal{B})}^{0}\rangle$ their two-fold degenerate
parental eigenstates, correspondingly. Here index
$\mathcal{A}$ stands for the product state, $\vert
\psi_{\mathcal{A}}^{0}\rangle$=$\vert \phi_{k}\rangle \otimes
\vert \phi_{l}\rangle$, whereof the larger part of the energy, 
$E_\mathcal{A}^0=\varepsilon_k^A+\varepsilon_l^B$, is
located in the system $A$, i.e. $\epsilon^{A}=\epsilon_{k}$,
$\epsilon^{B}=\epsilon_{l}$, $\epsilon_{k} > \epsilon_{l}$. The
eigenstate $\vert \psi_{\mathcal{B}}^{0}\rangle$ is given by the
permutation, $\vert \phi_{l}\rangle \otimes \vert \phi_{k}\rangle
$. In the limit of the arithmetic-mean equilibration the
tridiagonal structure of the transformation matrix allows for
energy exchange between levels with the same energies,
$\epsilon^{A}=\epsilon^{B}$, only. A strengthening of the
interaction violates this `level-to-level'  rule, so that more
than two eigenstates of the composite system can exchange their
energies. Yet the interaction of any strength preserves the permutation
symmetry. Therefore, a step beyond the arithmetic-mean
equilibration consists in inclusion of a `coupling' between the
pair of states $\vert \psi_{s, a}^{\lambda}\rangle$ and its
neighbor (closest in total energy) pair, $\vert \tilde{\psi}_{s,
a}^{\lambda}\rangle$. In terms of the transformation matrix elements 
the coupling means that the pair $\vert \psi_{\mathcal{A},
\mathcal{B}}^{0}\rangle$ contributes to the states $\vert
\tilde{\psi}_{s, a}^{\lambda}\rangle$, and the pair $\vert
\tilde{\psi}_{\mathcal{A}, \mathcal{B}}^{0}\rangle$ contributes to
the states $\vert \psi_{s, a}^{\lambda}\rangle$. The block corresponding
to this unitary transformation has the following form
\begin{equation}
\left(
 \begin{array}{cccc}
    \vert \psi_{s}^{\lambda}\rangle  \\
    \vert \psi_{as}^{\lambda}\rangle  \\
    \vert \tilde{\psi}_{s}^{\lambda}\rangle  \\
        \vert \tilde{\psi}_{as}^{\lambda}\rangle
\end{array}\right) =     \left(
\begin{array}{cccc}
    a & a & b &  b \\
    c &   - c & d & -d \\
    e &  e &   f &  f \\
    g & -g & h &   - h
\end{array}
\right)
\left(
\begin{array}{cccc}
    \vert \psi_{\mathcal{A}}^{0}\rangle  \\
    \vert \psi_{\mathcal{B}}^{0}\rangle  \\
    \vert \tilde{\psi}_{\mathcal{A}}^{0}\rangle  \\
    \vert \tilde{\psi}_{\mathcal{B}}^{0}\rangle
\end{array}\right).
\label{eq:Xsys}
\end{equation}
Assuming that only one $4 \times 4$ block is present in the transformation matrix, after tracing over one system and neglecting the oscillating elements we obtain the following `equilibrium' populations of the energy levels:
\begin{eqnarray}
\label{eq:equlibrium}
p_k^{ A} = p_k^{ B} = \frac{1}{2}\left[p_k^{ A}(0)+p_k^{ B}(0)\right] - \delta P ,\nonumber \\
p_l^{ A} = p_l^{ B} = \frac{1}{2}\left[p_l^{ A}(0)+p_l^{ B}(0)\right] - \delta P, \nonumber \\
p_{\tilde{k}}^{ A} = p_{\tilde{k}}^{ B} = \frac{1}{2}\left[p_{\tilde{k}}^{ A}(0)+p_{\tilde{k}}^{ B}(0)\right] + \delta P, \nonumber \\
p_{\tilde{l}}^{ A} = p_{\tilde{l}}^{ B} = \frac{1}{2}\left[p_{\tilde{l}}^{ A}(0)+p_{\tilde{l}}^{ B}(0)\right] + \delta P,
\end{eqnarray}
where $\delta P = \alpha (p_{\tilde{k}}^{ A}(0)
p_{\tilde{l}}^{ B}(0)+ p_{\tilde{l}}^{ A}(0)
p_{\tilde{k}}^{ B}(0)-p_{k}^{ A}(0) p_{l}^{ B}(0)-
p_{l}^{ A}(0) p_{k}^{ B}(0))$. The only parameter is
$\alpha=a^4+c^4+e^4+g^4-1/2$ $=b^4+d^4+f^4+h^4-1/2$, $-1/2 <
\alpha < 0$. Therefore, the $4 \times 4$ block produces the
unidirectional energy exchange between the pairs $\vert \psi_{s,
a}^{\lambda}\rangle$ and $\vert \tilde{\psi}_{s,
a}^{\lambda}\rangle$.

The above result can be generalized to the case of a $2M \times 2M$
block, which involves the energy exchange between $M$ pairs of
eigenstates. The energy exchange is then parameterized by $K$ parameters
$\alpha_{q}$, $q=1,..,K$, where $K$ is given by the number of ways to choose 
a pair from $M$ elements, i.e. $K=M(M-1)/2$.

When the transformation matrix has a multi-block structure,
a given single system eigenstate $\vert \phi_{k}\rangle$ may enter
several blocks. Then the quasi-equilibrium energy level population for the
state $| \phi_{k}\rangle$ is given by
\begin{eqnarray}
\label{eq:equl_general}
p_k^{A,B} \simeq \frac{1}{2}\left[p_k^{ A}(0)+p_k^{ B}(0)\right] +
\sum_{\{s\}_k}\sum_{q=1}^{K_s} \Phi_{kl_s,k_q l_q}\alpha_q^s,~~~~~~
\end{eqnarray}
where the sum index, $s$, runs over the numbers of those blocks $\{s\}_k$ that the $k$-th state
participates in, $K_s=M_s(M_s-1)/2$ is the number of parameters $\alpha_q^s$ in the
$s$-th block of size $2M_s \times 2M_s$, $l_s$ is the index of a partner state
for the $s$-th block, and $k_q l_q$ denotes the product state
which exchanges the energy with the product state $kl_s$. The
typical multi-block structure of the transformation matrix
$\Lambda$ is depicted in Fig.~\ref{fig1s}(b).

To demonstrate that even a small number of $2M\times2M$ blocks
with $M>1$ in the transformation matrix $\Lambda$, ensures that the energy level 
populations, Eq.~(\ref{eq:equl_general}), approach the
Gibbs-like equilibrium, we use a simple model system with
$\mathcal{N}_S=6$ energy levels. We take the transformation
matrix with the structure displayed in Fig.~\ref{fig1s}(c).
According to (\ref{eq:equl_general}), the latter is fully
described by the seven parameters: $\alpha_1^1$ (1st $2 \times 2$ block),
$\alpha_1^2$ (2nd $2 \times 2$ block), $\alpha_1^3$, $\alpha_2^3$, $\alpha_3^3$
($3 \times 3$ block), $\alpha_1^4$ (4th $2 \times 2$ block) and $\alpha_1^5$
(5th $2 \times 2$ block). In the following, we rename, for simplicity, these seven
parameters again to $\alpha_i,\,i=1,\dots,7$. The $\mathcal{N} \times \mathcal{N}$ matrix,
with $\mathcal{N} = \mathcal{N}_S \times \mathcal{N}_S =36$ entries, describes the
composite system. The seven corresponding `exchange' terms,
$\Phi$, composed according to Eq.~(\ref{eq:equl_general}),
have the following forms:
\begin{eqnarray}
\label{eq:seven_p_model}
\alpha_1,~\Phi_{14,23} &=& p^{ A}_1(0)p^{ B}_4(0)+p^{ A}_4(0)p^{ B}_1(0)-p^{ A}_2(0)p^{ B}_3(0) \nonumber\\
        &&-p^{ A}_3(0)p^{ B}_2(0),\nonumber\\
\alpha_2,~\Phi_{15,24} &=& p^{ A}_1(0)p^{ B}_5(0)+p^{ A}_5(0)p^{ B}_1(0)-p^{ A}_2(0)p^{ B}_4(0)\nonumber\\
        &&-p^{ A}_4(0)p^{ B}_2(0),\nonumber\\
\alpha_3,~\Phi_{16,25} &=& p^{ A}_1(0)p^{ B}_6(0)+p^{ A}_6(0)p^{ B}_1(0)-p^{ A}_2(0)p^{ B}_5(0)\nonumber\\
        &&-p^{ A}_5(0)p^{ B}_2(0),\nonumber\\
\alpha_4,~\Phi_{16,34} &=& p^{ A}_1(0)p^{ B}_6(0)+p^{ A}_6(0)p^{ B}_1(0)-p^{ A}_3(0)p^{ B}_4(0)\nonumber\\
        &&-p^{ A}_4(0)p^{ B}_3(0),\nonumber\\
\alpha_5,~\Phi_{25,34} &=& p^{ A}_2(0)p^{ B}_5(0)+p^{ A}_5(0)p^{ B}_2(0)-p^{ A}_3(0)p^{ B}_4(0)\nonumber\\
        &&-p^{ A}_4(0)p^{ B}_3(0),\nonumber\\
\alpha_6,~\Phi_{26,35} &=& p^{ A}_2(0)p^{ B}_6(0)+p^{ A}_6(0)p^{ B}_2(0)-p^{ A}_3(0)p^{ B}_5(0)\nonumber\\
        &&-p^{ A}_5(0)p^{ B}_3(0),\nonumber\\
\alpha_7,~\Phi_{36,45} &=& p^{ A}_3(0)p^{ B}_6(0)+p^{ A}_6(0)p^{ B}_3(0)-p^{ A}_4(0)p^{ B}_5(0)\nonumber\\
        &&-p^{ A}_5(0)p^{ B}_4(0)\nonumber.
\end{eqnarray}
so the energy level populations, $p_k^{ A,B}(0), k=1,\dots,6$, approach and fluctuate around the equilibrium populations $p_k^{ A,B}$:
\begin{eqnarray}
\label{eq:seven_p_eq}
p_1^{ A,B} = p_1^{ar}+\alpha_1 \Phi_{14,23}+\alpha_2 \Phi_{15,24}+\alpha_3 \Phi_{16,25}+\alpha_4 \Phi_{16,34},\nonumber\\
p_2^{ A,B} = p_2^{ar}-\alpha_1 \Phi_{14,23}-\alpha_2 \Phi_{15,24}
-\alpha_3 \Phi_{16,25} ~~~~~~~~~~~~~~~~\nonumber\\
+\alpha_5 \Phi_{25,34}+ \alpha_6 \Phi_{26,35},\nonumber\\
p_3^{ A,B} = p_3^{ar}-\alpha_1 \Phi_{14,23}-\alpha_4 \Phi_{16,34}
-\alpha_5 \Phi_{25,34} ~~~~~~~~~~~~~~~~ \nonumber\\
-\alpha_6 \Phi_{26,35}+\alpha_7 \Phi_{36,45},\nonumber\\
p_4^{ A,B} = p_4^{ar}+\alpha_1 \Phi_{14,23}-\alpha_2 \Phi_{15,24}
-\alpha_4 \Phi_{16,34} ~~~~~~~~~~~~~~~~\nonumber\\
-\alpha_5 \Phi_{25,34}+\alpha_7 \Phi_{36,45},\nonumber\\
p_5^{ A,B} = p_5^{ar}+\alpha_2 \Phi_{15,24}-\alpha_3 \Phi_{16,25}+\alpha_5 \Phi_{25,34} ~~~~~~~~~~~~~~~~\nonumber\\
-\alpha_6 \Phi_{26,35}-\alpha_7 \Phi_{36,45},\nonumber\\
p_6^{ A,B} = p_6^{ar}+\alpha_3 \Phi_{16,25}+\alpha_4 \Phi_{16,34}+\alpha_6 \Phi_{26,35}+\alpha_7 \Phi_{36,45}, \nonumber
\end{eqnarray}
where $p_k^{ar} = (p_k^{ A}+p_k^{ B})/2$ denote the arithmetic-mean.

\begin{figure}[t]
\includegraphics[width=0.45\textwidth]{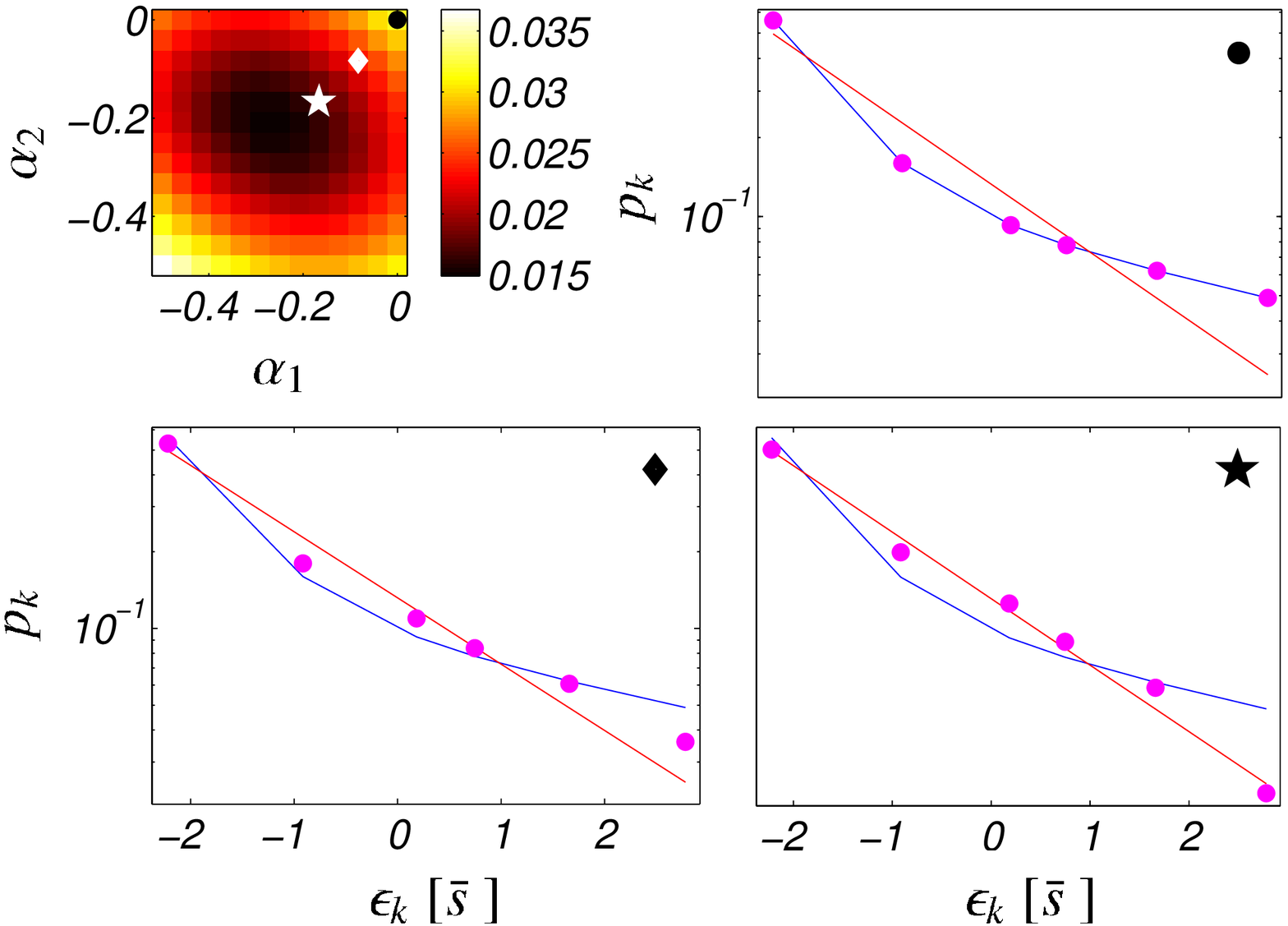}
\caption[Block-diagonal model] {{\bfseries Towards the canonical populations 
of energy levels.} The color diagram displays the root mean square
deviation, Eq.~(\ref{eq:delta_integr}), of the quasi-equilibrium energy 
level populations from the canonical populations, for the $6$-level model 
with the transformation matrix sketched in Fig.~\ref{fig1s}(c). 
The three subplots ($\bullet,\blacklozenge, \bigstar$) show the energy level 
populations (dots) in linear-log scale, for the three different sets of values of
block parameters, $\alpha_1=\alpha_j,\,j=2,\dots,7$, indicated by the
corresponding symbols on the color diagram. In addition, the canonical
populations (straight red lines) is plotted together with the
arithmetic-mean of initial populations (blue lines).} \label{fig2s}
\end{figure}

Figure~\ref{fig2s} displays the above equilibrium populations of the six energy
levels, $\epsilon_k,k=1,\dots,6$, possessed by the system of two bosons
in three-site lattice, with the initial temperatures
$k_\mathcal{B}T_{ A}=(\epsilon_2-\epsilon_1)/2$,
$k_\mathcal{B}T_{ B}= \epsilon_6-\epsilon_1$. To measure the
deviation from the canonical population, we calculate the root mean
square deviation (RMSD) of the populations $p_k^{ A}$ from the
canonical populations at the temperature $T_F$, evaluated from the
condition of energy conservation, Eq.~(7) of the main manuscript:
$\Delta(\alpha_1,\alpha_2,\dots,\alpha_7) =
\sqrt{\sum_{k=1}^{6}\left[p_k^{ A} - p_k(T_F)\right]^{2}/6}$.
We average the RMSD $\Delta(\alpha_1,...,\alpha_7)$ over the
region $\mathcal{P}$ with positive resulting populations of the energy levels,
$p_k^{ A,B} \geq 0$, in the parameter space,
$-0.5 < \alpha_j < 0, j=3,\dots,7$, and, in Fig.~\ref{fig2s},
plot the resulting two-variable function,
\begin{eqnarray}
\label{eq:delta_integr}
\tilde{\Delta} (\alpha_1,
\alpha_2)=\int_\mathcal{P} d\alpha_3 d\alpha_4 \dots d\alpha_7
\Delta(\alpha_1,\alpha_2,\dots,\alpha_7).
\end{eqnarray}
At the point $\alpha_1=\alpha_j=0$, with $j=2,\dots,7$ (indicated by the symbol $\bullet$,
in the color diagram) the transformation matrix is tridiagonal and the
energy level populations (dots) follow the arithmetic-mean of initial populations (solid blue lines), 
while any deviation from this point (e.g., at the two points $\alpha_1=\alpha_j$,
with $\alpha_1$ indicated by the symbols $\blacklozenge$ and $\bigstar$,
in the color diagram) drags the energy level populations towards the
canonical populations (straight red lines, in linear-log scale).

\section{The models of two finite quantum systems in contact}

The boson model is represented by the Bose-Hubbard Hamiltonian,
\begin{equation}
\label{hubbard}
H_S = -\frac{J}{2}\sum_{l=1}^L \left(a^\dagger_l a_{l+1} + a^\dagger_{l+1} a_l\right)+\frac{U}{2}\sum_{l=1}^L n_l(n_l-1),
\end{equation}
where $a^\dagger_l$ ($a_l$) is the bosonic creation (annihilation)
operator on site $l$, $n_l=a^\dagger_l a_l$ is the particle
number operator, and $L$ is the total number of lattice sites. The
parameters $J$ and $U$ are the hopping and the on-site
interaction strengths, respectively. The contact interaction
between the bosons from different systems takes place at the one
site only,  $l_A=L$ ($l_B=1$),
\begin{equation}
\label{hubbard_int}
H^{\rm int} = U^{\rm int} n_L^{ A} \otimes n_1^{ B},
\end{equation}
with the on-site interaction strength $U^{\rm int}$. In all our
calculations we set the ratio $J/U=7/3$, which is far from the
case of degenerate spectrum at $J=0,U\neq 0$ or $U=0,J\neq
0$. Finally, without loss of generality, we set $U_{\rm
int}=J+U$. The Hilbert space of the system with $N$ bosons and
$L$ lattice sites is spanned by $\mathcal{N_S}=(L+N-1)!/(L-1)!/N!$
basis states.

\begin{figure}[t]
\includegraphics[width=0.435\textwidth]{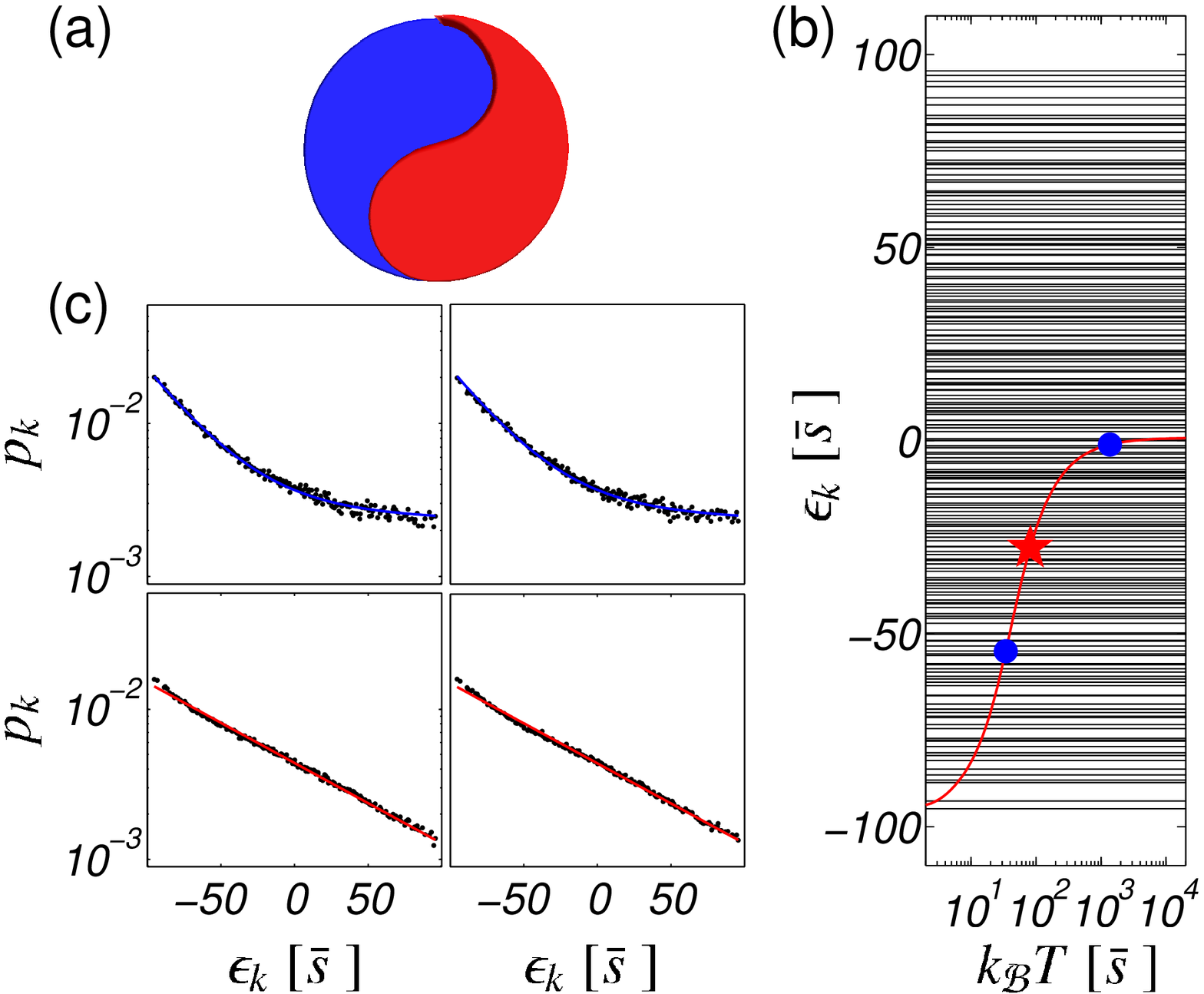}
\caption{{\bf Results for (a) a randomly synthesized GOE model of two quantum peers.} 
(b) Energy spectrum of a single system. The (red) line displays 
the dependence of the system mean energy, i.e., $E^S=\sum_{k} \epsilon_k
e^{-\epsilon_k/k_\mathcal{B} T}/Z_S$, on temperature $T$. The initial
temperatures of the `hot' system, $k_\mathcal{B}T_{ A}=1363.81~\bar{s}$,
and the `cold' system, $k_\mathcal{B}T_{ B}=34.09~\bar{s}$, are indicated
by the (blue) dots. The equilibrium temperature, $k_\mathcal{B}T_F=81.05~\bar{s}$,
calculated by using the total energy conservation, see Eq.~(7) of main manuscript,
is indicated by the (red) star. (c) Instantaneous `equilibrium' energy level
populations for the system $A$ (left column) and $B$ (right column),
in the regime of arithmetic-mean (top) and thermal (bottom) equilibration.
The exact arithmetic-mean of initial populations is depicted by the top (blue)
solid  lines, and the canonical populations for the temperature $T_F$
by the bottom (red) lines. The natural energy unit, $\bar{s}$, is given
by the mean energy level spacing of the single system,
$\bar{s}=(\epsilon_{\mathcal{N}_{S}}-\epsilon_1)/(\mathcal{N}_S-1)$.}
\label{fig4s}
\end{figure}

The parameters used in calculations: $N=L=5$, $J=13.29~\bar{s}$, $U=5.69~\bar{s}$,
$\lambda U^{\rm int}=0.19~\bar{s}$ (the arithmetic-mean equilibration, Fig.~1(c) of main manuscript, top) and
$\lambda U^{\rm int}=1.9~\bar{s}$ (the thermal equilibration, Fig.~1(c) of main manuscript, bottom),
$k_\mathcal{B}T_{ A}=5(U+J)=94.91~\bar{s}$, $k_\mathcal{B}T_{ B}=U+J=18.98~\bar{s}$, 
$k_\mathcal{B}T_F=33.92~\bar{s}$.

To synthesize the Hamiltonian of the second model, we employed
$\mathcal{N}_S\times \mathcal{N}_S$ matrix specimens from a
Gaussian Orthogonal Ensemble (GOE) \cite{matrix}. We use a
system with $\mathcal{N}_S=192$ states, so that the composite
system exhibits $\mathcal{N}=192^2=36,864$ energy levels. The matrix
elements of Hamiltonian $H_S$ were taken from the symmetric normal
distribution with dispersion $\sigma$, set to one in dimensionless
units. It has been rescaled then to the mean level spacing
$\bar{s}$, so that $\sigma = 3.55 \bar{s}$. The constraint on the
Hamiltonian, $[H_{S}]_{n,n'}=[H_{S}]_{n',n}$, $n,n'=1,...,N$,
provides its hermicity. The interaction Hamiltonian, $H^{\rm
int}$, was modelled by the direct product of two identical random
matrices from GOE, generated independently, by using the same
procedure. The parameters used in calculations are: $\lambda U^{\rm
int}=2.72 \cdot 10^{-4} ~\bar{s}$ (the arithmetic-mean equilibration,
Fig.~\ref{fig4s}(c), top) and $\lambda U^{\rm int}=5.45
\cdot 10^{-3} ~\bar{s}$ (the thermal equilibration, Fig.~\ref{fig4s}(c), bottom),
$k_\mathcal{B}T_{ A}=1363.81~\bar{s}$, $k_\mathcal{B}T_{ B}=34.09~\bar{s}$,
$k_\mathcal{B}T_F=81.05~\bar{s}$.

\section{Crossover from the arithmetic-mean to the thermal equilibration}

\begin{figure}[t]
\includegraphics[width=0.45\textwidth]{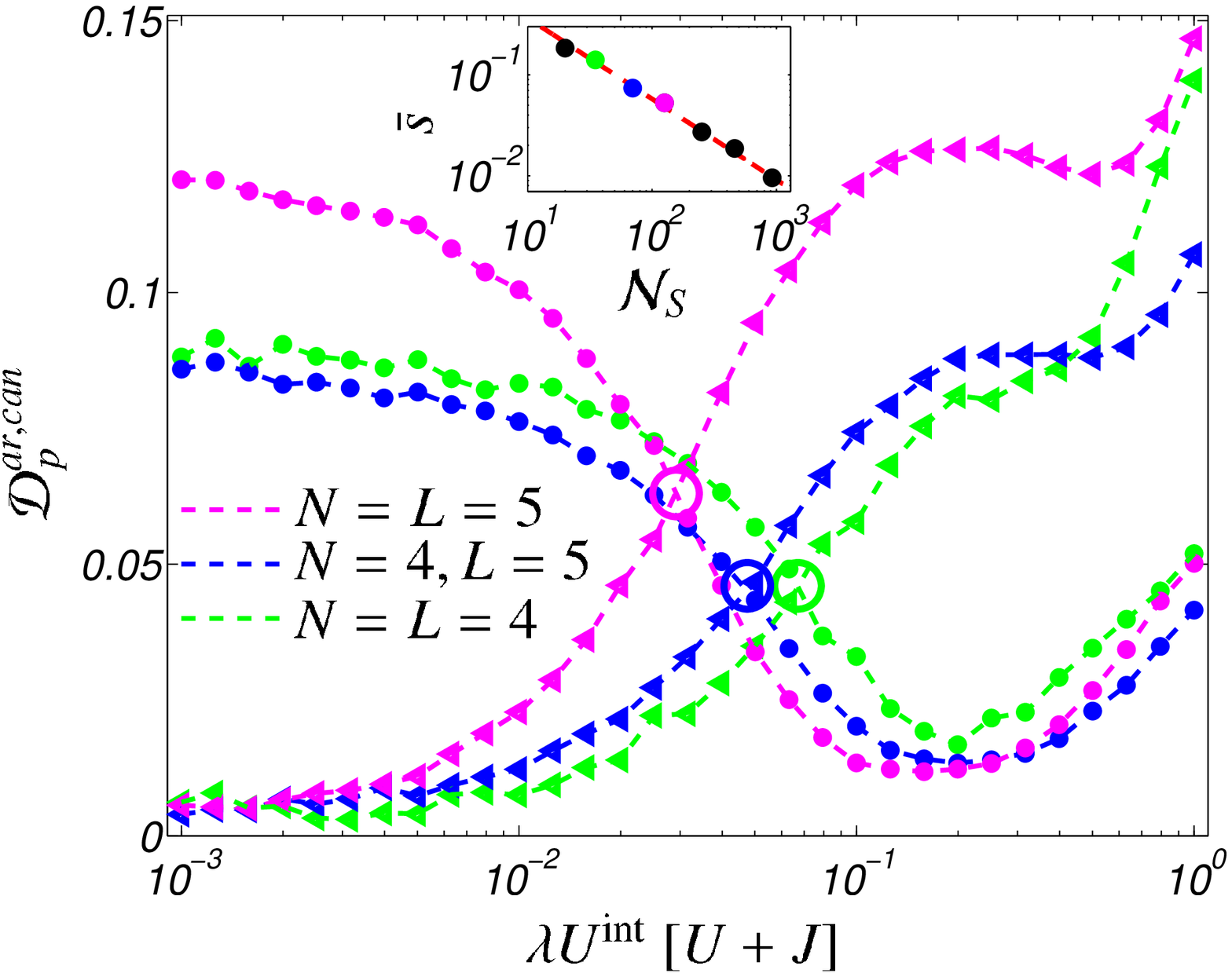}
\caption{{\bfseries Crossover from the arithmetic-mean to the thermal equilibration.}
The distance, Eq.~(\ref{trace_d}), between the time averaged energy level populations
of system $A$ after relaxation, and the arithmetic-mean (triangulares), and the thermal populations corresponding to the equilibrium energy (dots). The crossover is shown for the three different combinations of particle numbers, $N$, and lattice sizes, $L$. To guide the eye, identical markers are connected by dashed lines, whose crossing points 
are indicated by the open circles. The inset shows the dependence of the single system  mean level spacing, $\bar{s}=(\varepsilon_{\mathcal{N}_S}-\varepsilon_1)/(\mathcal{N}_S-1)$, Eq.~(\ref{hubbard}), on the size of the Hilbert space, $\mathcal{N}_S$, for the systems with $(N,L)$ equal to $(3,4)$, $(4,4)$, $(4,5)$, $(5,5)$, $(5,6)$, $(6,6)$, and $(6,7)$. The data points (dots) are fitted by a power-law, $\bar{s} \propto \mathcal{N}_S^{\xi}$ (dashed line), with the exponent $\xi \approx -0.8$.}
\label{fig5sa}
\end{figure}

In this section we provide the details of the crossover between the regimes of
`arithmetic-mean' and thermal equilibrations. We analyze the state of the subsystem $A$ represented
by its energy level populations $p_k^A(t)$, after the relaxation from initial Gibbs state. 
To quantify the deviation of the actual `equilibrium' state from the arithmetic-mean or Gibbs-like states, 
we employ the distance between energy level populations:
\begin{equation}
\label{trace_d}
\mathcal{D}_p^{ar,can} = \sum_k \vert p_k^{ar,can} -\langle p_k^A\rangle_t\vert
\end{equation}
where $\langle \dots \rangle_t$ denotes the time-average evaluated after the equilibration,
$\tau_{rec}>t>\tau_{rel}$, and $p_k^{ar}$ ($p_k^{can}$) are the populations of the corresponding arithmetic-mean (Gibbs-like) state.

Figure~\ref{fig5sa} shows the results of the calculations for the bosonic model,
Eq.~(\ref{hubbard}-\ref{hubbard_int}), with three different sets of the particle numbers, $N$,
and the lattice sizes, $L$: $(N,L)=(4,4),(4,5),(5,5)$, which result in the Hilbert spaces of size, 
$N_S=35,70,126$, respectively. The other parameters of the corresponding 
Hamiltonians are the same as in the previous section, i.e. $J/U=7/3$, $k_\mathcal{B}T_{ A}=5(U+J)$, and
$k_\mathcal{B}T_{ B}=U+J$.

The numerical results  demonstrate the presence of a smooth crossover between the arithmetic-mean and
thermal regimes of equilibration. The regimes reveal themselves by the plateau-like minima of $\mathcal{D}_p^{ar,can}$. The increase of system size $\mathcal{N}_S$ sharpens the crossover,
note the log-scale in Fig.~\ref{fig5sa}. In addition, there are three notable features: (i)~the crossover
region shifts to the region of smaller coupling strengths, thus squeezing the region of arithmetic-mean 
equilibration with increasing $\mathcal{N}_S$; (ii)~the region of thermal equilibration 
extends and the distance $\mathcal{D}^{can}_p$ decreases with increasing $\mathcal{N}_S$; (iii)~the 
finite region of thermal equilibration is followed by the regime where the the weak coupling 
condition, Eq.~(6) of main manuscript, is violated, causing the quasi-equilibrium energy level populations 
to deviate significantly from the thermal populations.

\begin{figure}[t]
\includegraphics[width=0.43\textwidth]{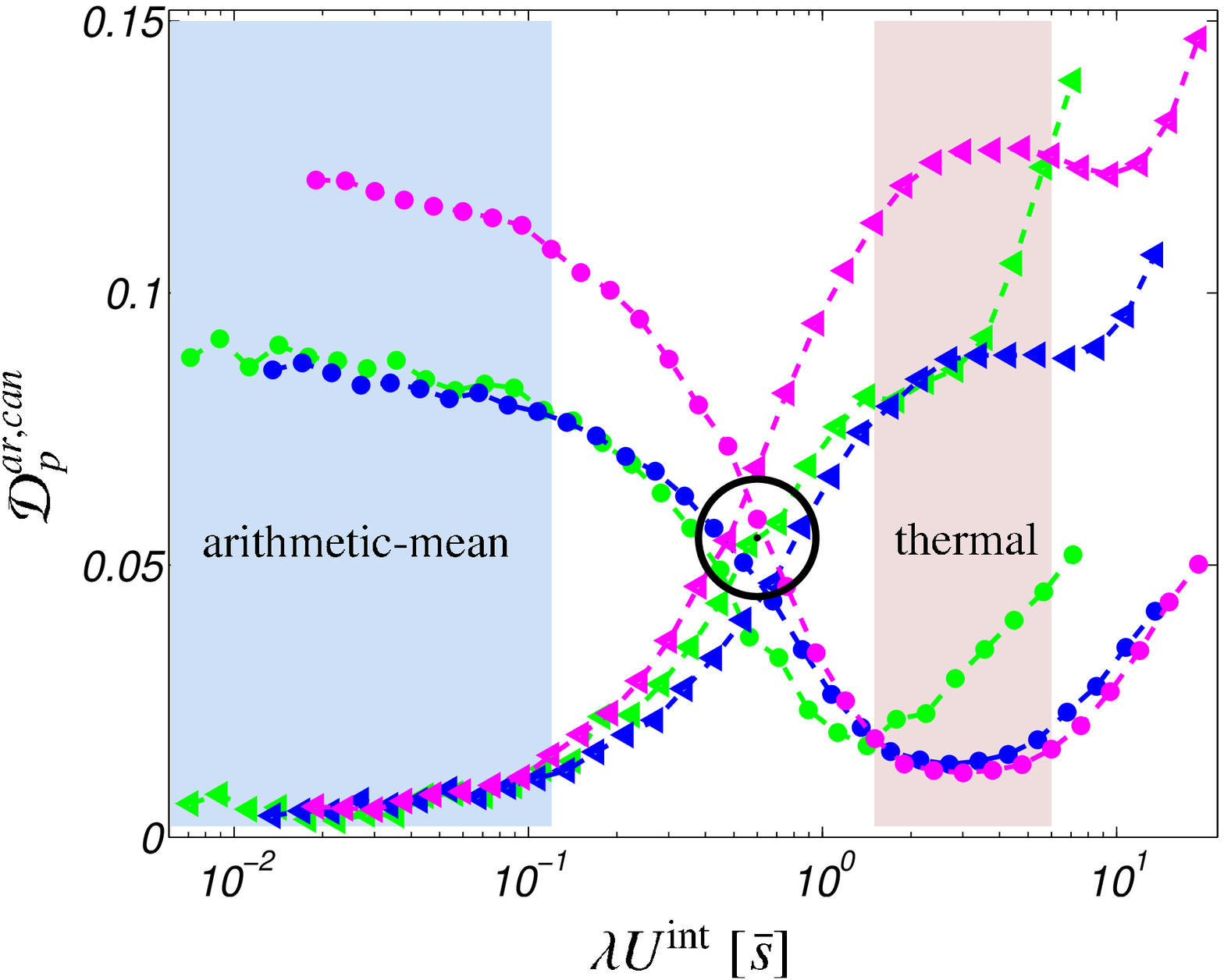}
\caption{{\bfseries The same distances as in Fig.~{\ref{fig5sa}}, but as functions of the coupling 
strength rescaled by the mean level spacing of the corresponding system.} After the scaling, in 
contrast to Fig.~{\ref{fig5sa}}, the systems of different size share the same crossover region (open circle). Note, that due
to the logarithmic scale of $x$-axis, the region of the thermal equilibration is much broader 
than that of the arithmetic-mean equilibration.}
\label{fig5sb}
\end{figure}

The density of states for the bosonic model, Eq.~(\ref{hubbard}), with fixed Hamiltonian parameters 
and the mean occupation number $\langle n \rangle=N/L\sim 1$, possesses essentially identical shapes. 
Thus we expect that the mean energy level spacing, $\bar{s}=(\varepsilon_{\mathcal{N}_S}-\varepsilon_1)/(\mathcal{N}_S-1)$,
sets the proper energy scale for the crossover region between the arithmetic-mean
and thermal equilibrations for different system sizes, $\mathcal{N}_S$.

The results of the scaling depicted in Figure~\ref{fig5sb} support this hypothesis.
The inset of Fig.~\ref{fig5sa} shows that the mean energy level spacing  scales like
$N_S^{\xi}$, with the exponent $\xi\sim-0.8$. This also provides us with the
scaling of the arithmetic-mean region with increasing of the system size. The scaling exponent seems
to depend on the mean lattice occupation number $\langle n \rangle$ and the ratio of the Hamiltonian
parameters, $J/U$, only.

\begin{figure}[t]
\includegraphics[width=0.45\textwidth]{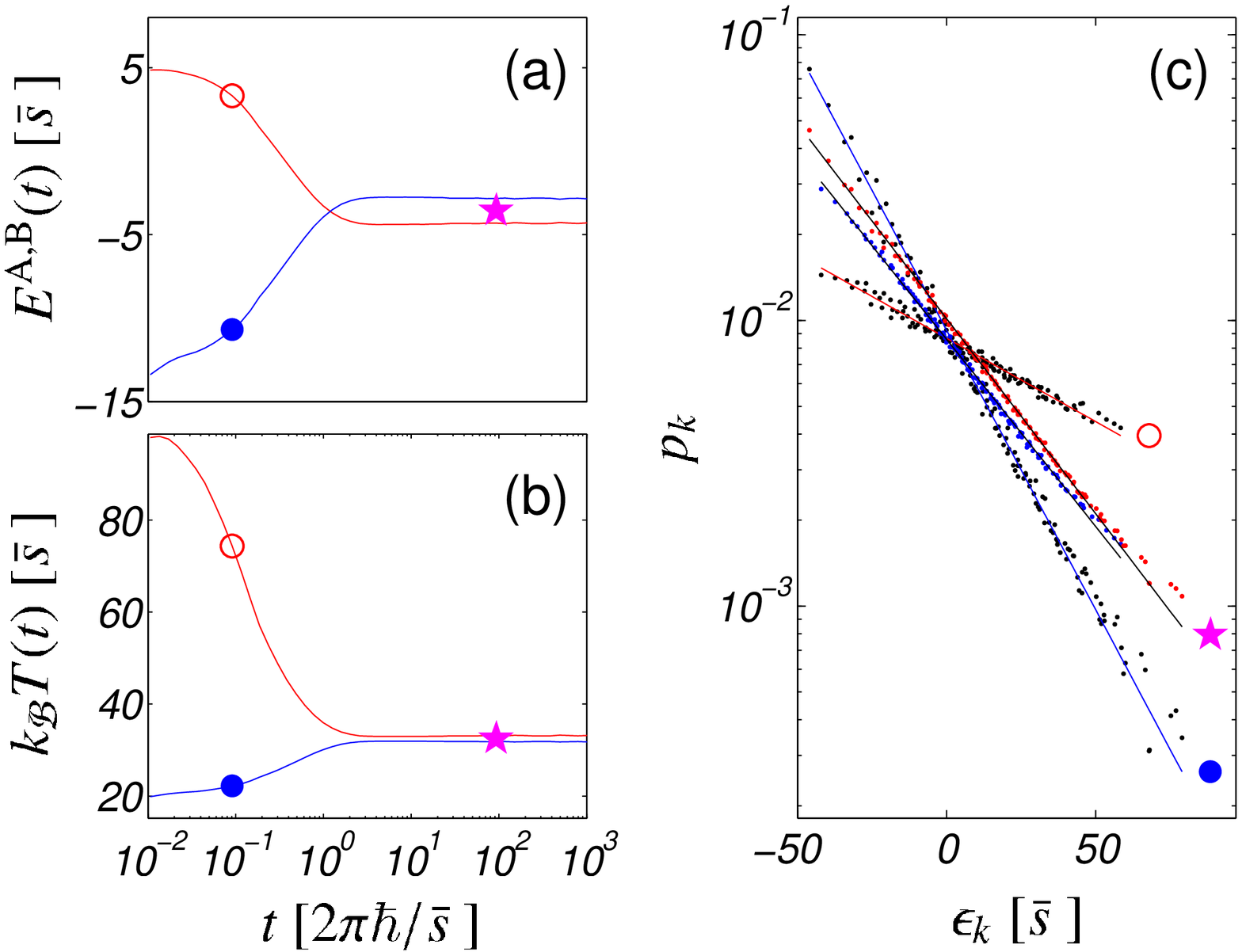}
\caption{{\bfseries Relaxation pathways for the bosonic model with two different systems in contact.}
Both systems are initially prepared in Gibbs states at different temperatures.
(a) The evolution of the mean system energies and (b) the corresponding temperatures of the `hot' system $A$
(with $N=4$ bosons on the lattice with $L=6$ sites) and the `cold' system $B$ (with $N=L=5$)
are shown by the red and blue lines, respectively. The time is plotted in units of inverse mean
level spacing $\bar{s}$ of the system $B$. (c) The energy level populations of both the systems
are displayed at different moments of time, marked by the corresponding symbols in (a, b). The solid
line corresponds to the Gibbs energy level populations at the temperatures evaluated from 
the temporal values of mean system energies. The Hamiltonian parameters for both the system and the initial system temperatures
are the same as in the case of the two identical systems in main manuscript, i.e. $T_A=5(U+J)=94.91~\bar{s}$ and 
$T_B=U+J=18.98~\bar{s}$, while the coupling constant only is five times larger, 
$\lambda U^{\rm int}=9.5~\bar{s}$.} \label{fig3s}
\end{figure}

\section{Different systems in contact}

Here we present the example of the equilibration process between two quantum systems with different spectra and
initial temperatures, but the same size of the Hilbert space, $\mathcal{N}_A=\mathcal{N}_B$, see Fig.~\ref{fig3s}. 
The lack of the permutation symmetry, $A\leftrightarrow B$, leads to the absence of two-fold degeneracies in the 
composite system at $\lambda=0$, meaning that an infinitesimal coupling strength $\lambda>0$ does not modify 
significantly the spectrum of composite system any more, in contrast to the previous case of two identical 
systems in contact. Therefore different subsystems do not exhibit the regime of arithmetic-mean equilibration. 
At very weak coupling, which would induce the arithmetic-mean equilibration for two identical systems, the 
`thermal equilibration' between different systems occurs locally, i.e. within a number of independent clusters of 
energy levels only. In order to render the thermal equilibration process between the subsystems of rather small sizes, 
$\mathcal{N}_A=\mathcal{N}_B=126$, we had to choose a relatively large value of the coupling constant, which is 
close to the upper bound of the weak coupling limit, Eq.~(6) of main manuscript.